\PassOptionsToPackage{table}{xcolor} 

\documentclass[acmtog,authorversion]{acmart}
\acmSubmissionID{1617}

\usepackage{booktabs} 
\usepackage{wrapfig} 

\usepackage[ruled]{algorithm2e} 

\SetAlFnt{\small}
\SetAlCapFnt{\small}
\SetAlCapNameFnt{\small}
\SetAlCapHSkip{0pt}

\acmJournal{TOG}

\begin{document}

\title[StructuredField]{StructuredField: Unifying Structured Geometry and Radiance Field}

\author{Kaiwen Song}
\email{sa21001046@mail.ustc.edu.cn}
\affiliation{%
  \institution{University of Science and Technology of China}
  \city{Hefei}
  \state{Anhui}  
  \postcode{230026}
  \country{China}
}

\author{Jinkai Cui}
\email{cuijk@mail.ustc.edu.cn}
\affiliation{%
  \institution{University of Science and Technology of China}
  \city{Hefei}
  \state{Anhui}
  \postcode{230026}
  \country{China}
}

\author{Zherui Qiu}
\email{zrqiu@mail.ustc.edu.cn}
\affiliation{%
  \institution{University of Science and Technology of China}
  \city{Hefei}
  \state{Anhui}
  \postcode{230026}
  \country{China}
}

\author{Juyong Zhang}
\email{juyong@ustc.edu.cn}
\affiliation{%
  \institution{University of Science and Technology of China}
  \city{Hefei}
  \state{Anhui}
  \postcode{230026}
  \country{China}
}

\renewcommand{\shortauthors}{Song et al.}
\newcommand{\mh}[1]{\multicolumn{1}{c}{\small #1}}
\newcommand{\mhs}[1]{\multicolumn{1}{c}{\small \shortstack[c]{#1}}} 

\begin{abstract}
Recent point-based differentiable rendering techniques have achieved significant success in high-fidelity reconstruction and fast rendering. However, due to the unstructured nature of point-based representations, they are difficult to apply to modern graphics pipelines designed for structured meshes, as well as to a variety of simulation and editing algorithms that work well with structured mesh representations. To this end, we propose StructuredField, a novel representation that achieves both a structured geometric representation of the reconstructed object and high-fidelity rendering reconstruction. We employ structured tetrahedral meshes to represent the reconstructed object. We reparameterize the geometric attributes of these tetrahedra into the parameters of 3D Gaussian primitives, thereby enabling differentiable, high-fidelity rendering directly from the mesh. Furthermore, a hierarchical implicit subdivision strategy is utilized to ensure a conformal mesh structure while empowering the representation to capture multi-scale details. To maintain geometric integrity during optimization, we propose a novel inversion-free homeomorphism that constrains the tetrahedral mesh, guaranteeing it remains both inversion-free and self-intersection-free during the optimization process and in the final result. Based on our proposed StructuredField, we achieve high-quality structured meshes that are completely inversion-free and conformal, while also attaining reconstruction results comparable to those of 3DGS. We also demonstrate the applicability of our representation to various applications such as physical simulation, deformation, and level-of-detail. Code available at \url{https://github.com/kevin2000A/StructuredField}.
\end{abstract}



\begin{teaserfigure}
  \includegraphics[width=\textwidth]{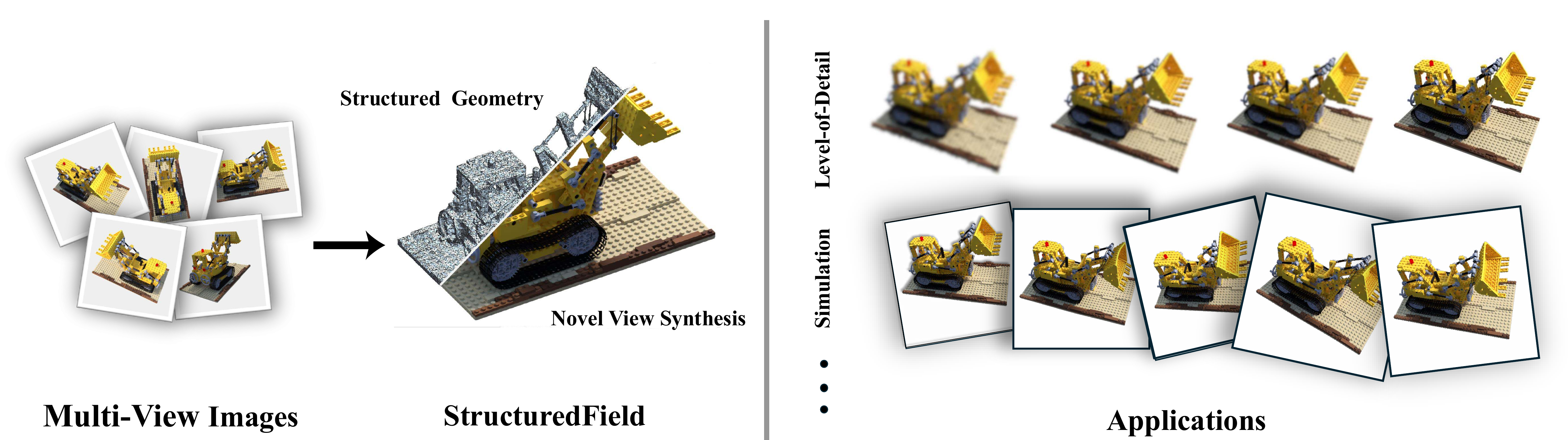}
  \caption{\textbf{StructuredField} represents and reconstructs scene using a structured tetrahedral mesh. This novel structured 3D representation enables a variety of applications, including physical simulations, deformations, and more.}
  \Description{StructuredField represents and reconstruct scene using a structured tetrahedral mesh. This novel structured 3D representation enables a variety of applications, including physical simulations, deformations, level-of-detail and more.}
  \label{fig:teaser}
\end{teaserfigure}

\maketitle


\section{Introduction}

Photorealistic and real-time rendering of 3D scenes is a central pursuit in computer graphics, both in academic research and practical applications. Traditional 3D representations, such as meshes~\cite{botsch2005high,liu2019soft,hoppe2023progressive} and point clouds~\cite{rusinkiewicz2000qsplat,yifan2019differentiable}, facilitate real-time rendering through rasterization techniques that align with modern GPU rendering architectures. Nevertheless, these methods frequently produce low-quality rendering, exhibiting missing geometric details and blurry artifacts. In contrast, emerging differentiable volumetric rendering methods~\cite{barron2021mip,muller2022instant,chen2022tensorf}, including neural radiance fields~\cite{mildenhall2021nerf}, demonstrate the capability to reconstruct 3D scenes in an end-to-end manner using multi-view images, thus achieving high fidelity and maintaining intricate details. However, the reliance on ray tracing-based rendering pipelines and the necessity for extensive sampling points pose challenges to rapid rendering in NeRF variants. 

Recently, advancements in point-based differentiable rendering techniques~\cite{kerbl3Dgaussians,held20243d,huang2024deformable,Huang2DGS2024} have demonstrated notable success in enhancing both rendering speed and high-fidelity reconstruction by leveraging smooth primitives within rasterization-based pipelines. While these point-based rendering techniques offer numerous advantages, they inherently represent scenes as unstructured point clouds. In contrast, contemporary graphics pipelines are predominantly designed for structured representations, such as polygonal meshes. These structured forms are widely adopted in a variety of applications, including animation~\cite{james2005skinning,sorkine2007rigid}, physical simulation~\cite{ando2013highly,baraff2023large,liu2013fast}, and editing~\cite{yu2004mesh,zorin1997interactive}, where an explicit and coherent geometric structure is essential. Consequently, the absence of such inherent object structure in point-based methods significantly constrains their broader applicability.

Our primary aim in this study is to bridge the gap between recent high-fidelity differentiable point-based representations and traditional graphics pipelines tailored for structured meshes. To this end, we introduce StructuredField, a novel method that unifies high-fidelity differentiable rendering and high-quality structured geometry. Our fundamental representation within StructuredField is the tetrahedral mesh, as it is suited for volumetric data and has been widely utilized in volume rendering~\cite{ma1997scalable,yagel1996hardware,weiler2003hardware}.  To enable high-fidelity visual reconstruction, StructuredField reparameterizes the attributes of 3D Gaussian primitives with the geometric parameters of the underlying tetrahedral mesh. This strategy facilitates the direct, end-to-end differentiable optimization of the tetrahedral mesh using image-based losses, thereby achieving detailed reconstruction and high-quality rendering.

However, directly optimizing the vertex positions of tetrahedral meshes can readily lead to anomalous geometric structures, such as self-intersections or element inversions. To prevent these issues and ensure the integrity of the geometric structure throughout the optimization process, we analyze two main causes of mesh anomalies: self-intersections and element inversions, and propose a novel orientation-preserving homeomorphism~\cite{dinh2016density,behrmann2019invertible}, which constrains the feasible deformation space of the mesh vertices, strictly maintaining a valid, intersection-free, and inversion-free tetrahedral mesh.

Furthermore, the capacity of a 3D representation to capture multi-scale details, like multi-resolution hash encodings in Instant-NGP~\cite{muller2022instant} and adaptive density control in 3D Gaussian Splatting~\cite{kerbl3Dgaussians}, is fundamental for achieving high-fidelity reconstruction. However, when extending to tetrahedral representations, achieving operations similar to adaptive density control often requires explicit, multi-level subdivision of the tetrahedra. Such explicit subdivision inherently introduces non-conformal interfaces between tetrahedral elements, which are problematic for downstream applications like physical simulation. Instead, we utilize an implicit subdivision strategy that adaptively refines regions for rendering by introducing implicit child tetrahedra without modifying the fixed topology of the initial conformal mesh. Consequently, by applying these two strategies, our representation inherently ensures high-quality geometry for simulation while providing sufficient expressive power for high-fidelity reconstruction.

Through comprehensive experimentation, we demonstrate that our approach achieves rendering quality comparable to or even exceeding that of recent point-based rendering techniques. Furthermore, our representation can be seamlessly integrated into various applications like physics simulation and deformation without necessitating algorithmic redesigns or extensive modifications. Our primary contributions can be summarized as:
\begin{itemize}
\item We introduce StructuredField, a novel 3D representation that seamlessly combines structured tetrahedral geometry with high-fidelity differentiable rendering capabilities through the reparameterization of 3D Gaussian functions.
\item We design a novel orientation-preserving homeomorphism constraint to ensure that the tetrahedral mesh remains inversion-free and self-intersection-free during optimization.
\item We propose an implicit subdivision strategy for tetrahedral meshes that maintains a high-quality base mesh structure suitable for physical simulations while ensuring adaptive multi-scale details for high-fidelity reconstructions.
\end{itemize}

\begin{figure*}[htbp]
    \centering
    \includegraphics[width=1.0\linewidth]{fig/newpipeline.pdf}
    \caption{\textbf{Overview of StructuredField.} Given multi-view images as input, we reconstruct the 3D scene using a structured tetrahedral mesh. Our implicit multi-level subdivision module iteratively refines the initial input tetrahedra, creating a hierarchical structure. The root nodes of this hierarchy (the base tetrahedra) form a high-quality geometric foundation; their vertex positions are optimized using an orientation-preserving homeomorphism to maintain an inversion-free and robustly structured geometry, which supports subsequent applications. The leaf node tetrahedra, capturing the finest adaptive details, are then reparameterized into 3D Gaussian Splatting (3DGS) primitives for high-fidelity differentiable rendering. Based on our structured mesh representation, the reconstructed model can be further applied to physical simulations, deformation, and other graphics applications.}
    \label{fig:pipeline}
\end{figure*}

\section{Related work}
\subsection{3D Reconstruction}
Reconstructing 3D scenes from multi-view images is a longstanding problem in both computer graphics and computer vision. Traditional 3D reconstruction techniques include Structure-from-Motion (SfM) pipelines~\cite{sfm,sfm2} to estimate camera poses and obtain sparse point clouds, followed by surface reconstruction through dense multi-view stereo~\cite{mvs1,mvs2,mvs3}. These methods rely on hand-crafted features to acquire fine textures and geometry, and struggle to reconstruct view-dependent colors. Significant advancements have been achieved in NVS, particularly since the introduction of Neural Radiance Fields (NeRF)~\cite{mildenhall2021nerf}. The original NeRF represents the scene as an MLP, which maps positional encodings of spatial locations and directions to attributes including color and density, and utilizes a volume rendering process to achieve realistic rendering. Various works have enhanced the performance of NeRF~\cite{muller2022instant,barron2023zip,barron2021mip} or extended them to large scenes~\cite{barron2022mip,tancik2022block}. More recently, 3DGS~\cite{kerbl3Dgaussians} optimizes anisotropic 3D Gaussian primitives, demonstrating real-time photorealistic reconstruction results. This method has been rapidly extended to multiple domains~\cite{lin2024vastgaussian,keetha2024splatam,tang2023dreamgaussian,xiang2024flashavatar}. Despite these successes, point-based representations are unstructured, limiting their further applications. In this paper, we demonstrate detailed reconstruction while maintaining a structured tetrahedral mesh, and showcase its subsequent applications.

\subsection{Mesh-based Representation}
Explicit representations have served as a cornerstone within 3D modeling and computer graphics for decades~\cite{woo1999opengl}. Conventional geometric representations like point clouds, voxels and polygonal meshes have been extensively revisited in the context of 3D deep learning. Polygonal meshes are particularly attractive due to their structured geometry and efficient rendering properties. Recently, differentiable rendering methods~\cite{liu2019soft,pidhorskyi2025rasterized} leveraging mesh representations have enabled the production of high-quality renderings. However, the optimization processes inherent to mesh representations are often hampered by rigid topological constraints, resulting in limited flexibility and reduced capacity to accurately depict realistic appearances. Our method integrates inversion-free mapping to effectively regulate and maintain structural integrity during the mesh optimization process while ensuring high quality reconstruction, addressing these limitations.

On the other hand, NeRF~\cite{mildenhall2020nerf} has attracted considerable attention due to their ability to deliver high-fidelity renderings. Substantial research has focused on integrating implicit methods with mesh structures to extend and refine NeRF~\cite{wang2021neus}. For instance, some approaches~\cite{wang2023adaptive,wan2023learning,turki2024hybridnerf} utilize meshes to constrain the sampling regions within an adaptively learned narrow band of two explicit meshes. On the other hand, Tetra-NeRF~\cite{kulhanek2023tetra,rosu2023permutosdf} employs tetrahedral meshes as feature grids to accelerate the training process. Additionally, leveraging the fast rendering capabilities of meshes, Baked SDF and its variants~\cite{yariv2023bakedsdf,reiser2024binary} have baked radiance fields into meshes to facilitate real-time rendering. Mesh representations have also been pivotal in the deformation of radiance fields. NeRFShop and Cage-NeRF~\cite{peng2022cagenerf,jambon2023nerfshop} use meshes as cages to drive the deformation of radiance fields. 

Recently, point-based representations~\cite{kerbl3Dgaussians,huang2024deformable,Huang2DGS2024} have emerged, utilizing smooth primitives to represent radiance fields and employing rasterization techniques to accelerate rendering. In subsequent works, Mani-gs and GaMes~\cite{gao2024mani} employ high-quality triangle meshes or triangle soups as proxies, binding primitives to the triangle mesh to drive the deformation of 3D Gaussians.  Additionally, VR-GS~\cite{jiang2024vr} and D3GA~\cite{zielonka2023drivable} adopt tetrahedral meshes as proxies to facilitate the deformation of objects. However, these methods directly bind the mesh to Gaussians without optimizing the positions of the mesh vertices. As a result, any imperfections in the initial mesh can significantly degrade the final reconstruction quality. In contrast, our method reparameterizes Gaussians with tetrahedral mesh vertices and optimizes these vertices, thereby allowing for high-quality reconstruction.

\subsection{Inversion-free Mesh Optimization}
Maintaining mesh quality during the optimization process remains a challenging problem. Element inversion is a major factor causing mesh quality issues~\cite{fu2021inversion}. Traditional methods~\cite{schuller2013locally,rabinovich2017scalable} impose constraints on the Jacobian of the mapping during optimization to ensure the Jacobian remains positive, thereby enforcing that the mapping is orientation-preserving. In addition to being inversion-free, intersection-free boundaries are also necessary during the optimization. This is typically achieved by constraining the mapping to be bijective. Traditional methods~\cite{jiang2017simplicial,misztal2012topology} introduce scaffold meshes to convert the globally overlap-free constraint into a locally flip-free condition. Recently, invertible networks~\cite{behrmann2019invertible,dinh2016density} offer another solution, leading to their widespread adoption in recent 3D deformation tasks. For instance, NDR~\cite{cai2022neural} and Cadex~\cite{lei2022cadex} employ invertible networks to model the motion of objects, thereby achieving dynamic object reconstruction. NFGP~\cite{yang2021geometry} utilizes invertible networks to perform geometric processing on implicit surfaces. Compared to these methods, we propose a novel invertible network that implicitly and strictly constrains the mapping to be inversion-free and intersection-free, and apply it during the optimization process to guarantee the mesh quality.
\section{Method}


Given multi-view posed images of a scene, our primary objective is to reconstruct it with high fidelity and high-quality geometry. This enables seamless integration of the reconstructed 3D scene into existing computer graphics pipelines for applications such as rendering, animation, physical simulation, and deformation. To this end, we first introduce a novel tetrahedral mesh-based 3D representation that is structurally organized and facilitates differentiable optimization (Sec.~\ref{sec:diff_mesh}). The mesh is optimized using an orientation-preserving homeomorphism to ensure high-quality structured geometry (Sec.~\ref{sec:topo}). We further introduce a hierarchical implicit subdivision scheme to achieve conformal geometry while enabling multi-scale rendering detail (Sec. ~\ref{sec:conformal}). An overview of our representation and reconstruction pipeline is shown in Fig.~\ref{fig:pipeline}.

\subsection{Representation of StructuredField}
\label{sec:diff_mesh}

Given a tetrahedral mesh \(\mathcal{M} = (\mathcal{V}, \mathcal{T})\), where \(\mathcal{V}=\{v_i\}_{i=1}^N\) represents the set of vertices, \(\mathcal{T}=\{T_k\}_{k=1}^K\) is the set of tetrahedron. Each tetrahedron \(T_k \in \mathcal{T}\) is defined by four vertices \(\{v_{k_1}, v_{k_2}, v_{k_3}, v_{k_4}\}\). Tetrahedral meshes are generally challenging to render in a differentiable manner due to their discrete, non-smooth nature.

To address this limitation, we build upon recent point-based differentiable rendering techniques, such as 3D Gaussian Splatting (3DGS)~\cite{kerbl3Dgaussians}, 2D Gaussian Splatting (2DGS)~\cite{Huang2DGS2024}, and convex splatting~\cite{held20243d}. In our method, we establish a one-to-one correspondence between each tetrahedron and a primitive, using a differentiable reparameterization function  \(\mathbf{F}_{r}(\cdot)\) to map the parameters \(T_k\) of the tetrahedron to  the parameters \(\Theta_k\) of the primitive: 
\begin{equation}
    \Theta_{k} = \mathbf{F}_{r}\bigl(T_{k}\bigr).
    \label{eq:reparametrize}
\end{equation}

The gradients are propagated back to the parameters of tetrahedra through the differentiable reparameterization function, enabling the optimization of vertex positions and associated attributes in a fully differentiable framework. In the following, we demonstrate how the parameters of 3DGS can be reparameterized using the parameters of a tetrahedral mesh.


\begin{figure}[tbp]
    \centering
    \includegraphics[width=1.0\linewidth]{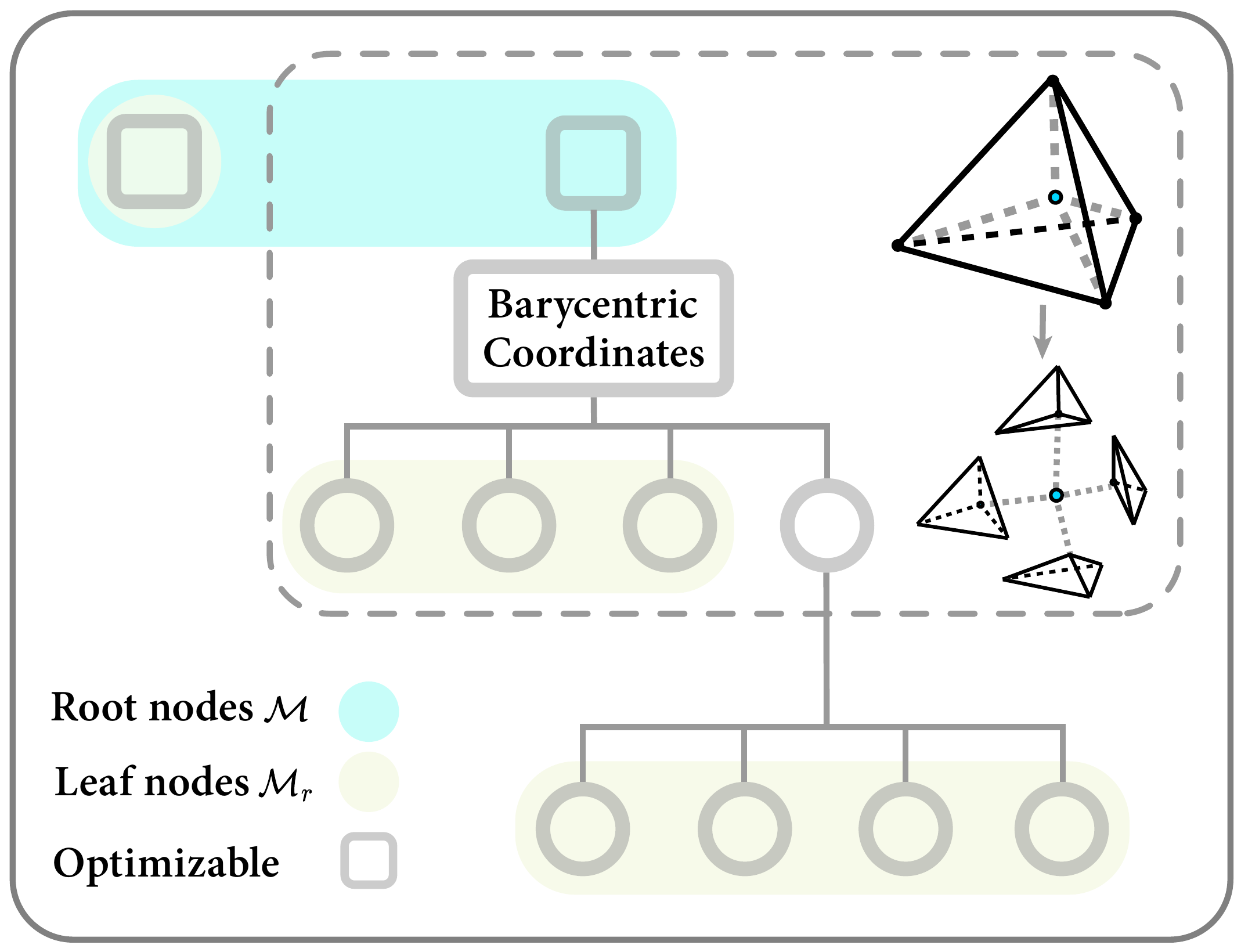}
    \caption{\textbf{Data Structure of Hierarchical Implicit Subdivision. }Base tetrahedra  can be iteratively subdivided in a 1-to-4 manner, forming a quadtree-like hierarchy. The root node tetrahedra constitute a high-quality and conformal geometry, designated for downstream applications such as physical simulation. Leaf node tetrahedra are utilized for high-quality rendering. The positions of the implicitly defined subdivision control points are optimized via their barycentric coordinates within their respective parent tetrahedra, while the positions of the root node vertices are optimized using our Orientation-preserving Homeomorphism. }
    \label{fig:tree}
\end{figure}

\paragraph{Reparametrize 3D Gaussians} 3DGS~\cite{kerbl3Dgaussians} represents a scene with a set of 3D Gaussians $\mathcal{G} = \{\boldsymbol{g}_{k}\}_{k=1}^{K}$, where each Gaussian $\boldsymbol{g}_{k}$ encodes the following attributes: mean $\boldsymbol{\mu}_{k} \in \mathbb{R}^3$, scales $\boldsymbol{s}_{k} \in \mathbb{R}^{3}$, rotation $\boldsymbol{r}_{k} \in \mathbb{R}^{4}$, color $\mathbf{c}_{k} \in \mathbb{R}^3$, and opacity $o_{k} \in \mathbb{R}$.  In our representation, each tetrahedron \(T_{k}\) contains four vertices with positions \(\{\mathbf{v}_{k_i}\}_{i=1}^4\), spherical harmonic coefficients \(\{\mathbf{SH}_{k_i}\}_{i=1}^4\), and weights \(\{w_{k_i}\}_{i=1}^4\). We first compute the mean \(\boldsymbol{\mu}_k\) and an initial, PCA-derived covariance matrix \(\boldsymbol{\Sigma}_{k}^{\prime}\) from the weighted vertex positions of \(T_{k}\), following a PCA-style approach~\cite{abdi2010principal}:
\begin{equation}
    \boldsymbol{\mu}_k = \frac{\sum_{i=1}^{4} w_{k_i} \, \mathbf{v}_{k_i}}{\sum_{i=1}^{4} w_{k_i}}, \quad
    \boldsymbol{\Sigma}_{k}^{\prime} = \sum_{i=1}^{4} w_{k_i} \bigl(\mathbf{v}_{k_i} - \boldsymbol{\mu}_k\bigr)\bigl(\mathbf{v}_{k_i} - \boldsymbol{\mu}_k\bigr)^\top.
    \label{eq:mean_pca_covariance}
\end{equation}
This \(\boldsymbol{\Sigma}_{k}^{\prime}\) captures the initial anisotropic shape implied by the vertex distribution of \(T_{k}\). The optimizable quaternion \(\mathbf{q}_{k}\) is converted to its corresponding rotation matrix \(\Delta\mathbf{R}_{k}\). The final anisotropic covariance matrix \(\boldsymbol{\Sigma}_k\) for the 3D Gaussian is then obtained by applying this learned rotation:
\begin{equation}
    \boldsymbol{\Sigma}_k = \Delta\mathbf{R}_{k} \boldsymbol{\Sigma}_{k}^{\prime} \Delta\mathbf{R}_{k}^T.
    \label{eq:final_covariance}
\end{equation}
This explicit, optimizable rotation allows the 3D Gaussian primitive to more flexibly orient itself to fit the local scene content associated with \(T_{k}\).

The opacity attribute $o_k$ is defined on the tetrahedron instead of on the vertices, so the opacity of its corresponding Gaussian is equivalent to the opacity of the tetrahedron. The color of the Gaussian is obtained by a weighted average of the colors computed from the spherical harmonic function of each vertex:


\begin{equation}
\mathbf{c} = \frac{\sum_{i=1}^4 w_i \mathbf{c}_i}{\sum_{i=1}^4 w_i}, \quad \mathbf{c}_i = \mathbf{SH}_i(\mathbf{d}),
\label{eq:color}
\end{equation}
where \(\mathbf{d}\) is the direction from the camera center to the vertex.

After reparameterizing 3D Gaussians with tetrahedral mesh, the scene can be rendered in a differentiable manner. However, freely optimizing the tetrahedral vertices leads to low-quality tetrahedral mesh, as shown in Fig.~\ref{fig:abla}. In Section~\ref{sec:topo}, we introduce the orientation-preserving homeomorphism to address this issue.

\subsection{Hierarchical Implicit subdivision}
\label{sec:conformal}

\begin{figure*}[htbp]
    \centering
    \includegraphics[width=\linewidth]{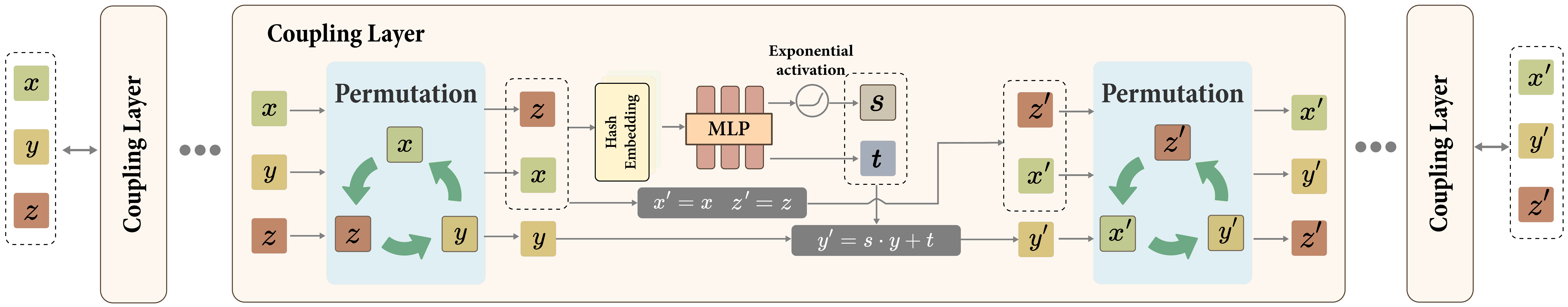}
    \caption{\textbf{Invertible neural network architecture.} By applying a permutation strategy, the input vector is divided into two contiguous parts, ensuring that the resulting Jacobian matrix is upper triangular with positive diagonal elements. This strategy guarantees that the determinant of the Jacobian is positive, thus ensuring that the mapping is inversion-free. }
    \label{fig:inn}
\end{figure*}

High-fidelity rendering often requires capturing details at multiple scales, a capability demonstrated by many successful 3D reconstruction methods. For example, Instant-NGP~\cite{muller2022instant} uses multi-resolution hash encodings, combining features from various grid resolutions to efficiently represent intricate, multi-scale details. Similarly, 3D Gaussian Splatting~\cite{kerbl3Dgaussians} employs adaptive density control, refining Gaussian primitives in under reconstructed regions to better capture fine-scale information. Therefore, to equip our tetrahedral representation with a comparable capability to capture details across diverse scales, a mechanism for refining the tetrahedral elements themselves is essential.

While explicitly subdividing tetrahedra in the base mesh $\mathcal{M}$ can introduce finer geometric details, this direct modification of the mesh topology often creates non-conformal interfaces or degenerate elements. Such non-conformality and the presence of degenerate or ill-conditioned elements are highly problematic for downstream applications like physical simulation, which demand well-structured, conformal meshes, and also invariably increase the complexity of the simulation mesh. To circumvent these drawbacks and equip our representation with the capability to capture details at different scales, we propose a hierarchical implicit subdivision strategy.

\paragraph{Hierarchical Implicit Subdivision} During optimization, base tetrahedra in $\mathcal{M}$ identified for refinement undergo 1-to-4 subdivision, initiating the formation of a tree hierarchy. For each subdivided tetrahedron, the split is achieved by introducing a new, optimizable control point conceptually located within it. The position of this control point is parameterized by its barycentric coordinates relative to its respective parent tetrahedron. These coordinates are optimized during training, ensuring the control point remains within the parent's volume while allowing for adaptive shaping of the resulting child tetrahedra. This subdivision process can be applied recursively if multiple levels of detail are desired, further extending this hierarchical structure.

This hierarchical structure fulfills a crucial dual role in our representation. The root nodes of this tree consistently form the conformal mesh $\mathcal{M}$ designated for physical simulation, preserving its fixed topology and structural integrity. Conversely, the leaf tetrahedra of this tree hierarchy, representing the finest level of adaptive detail, constitute the mesh  $\mathcal{M}_{r}$ utilized for high-fidelity rendering.

\subsection{Inversion-free Structured Geometry}
\label{sec:topo}

\begin{figure}[htbp]
    \centering
    \includegraphics[width=\linewidth]{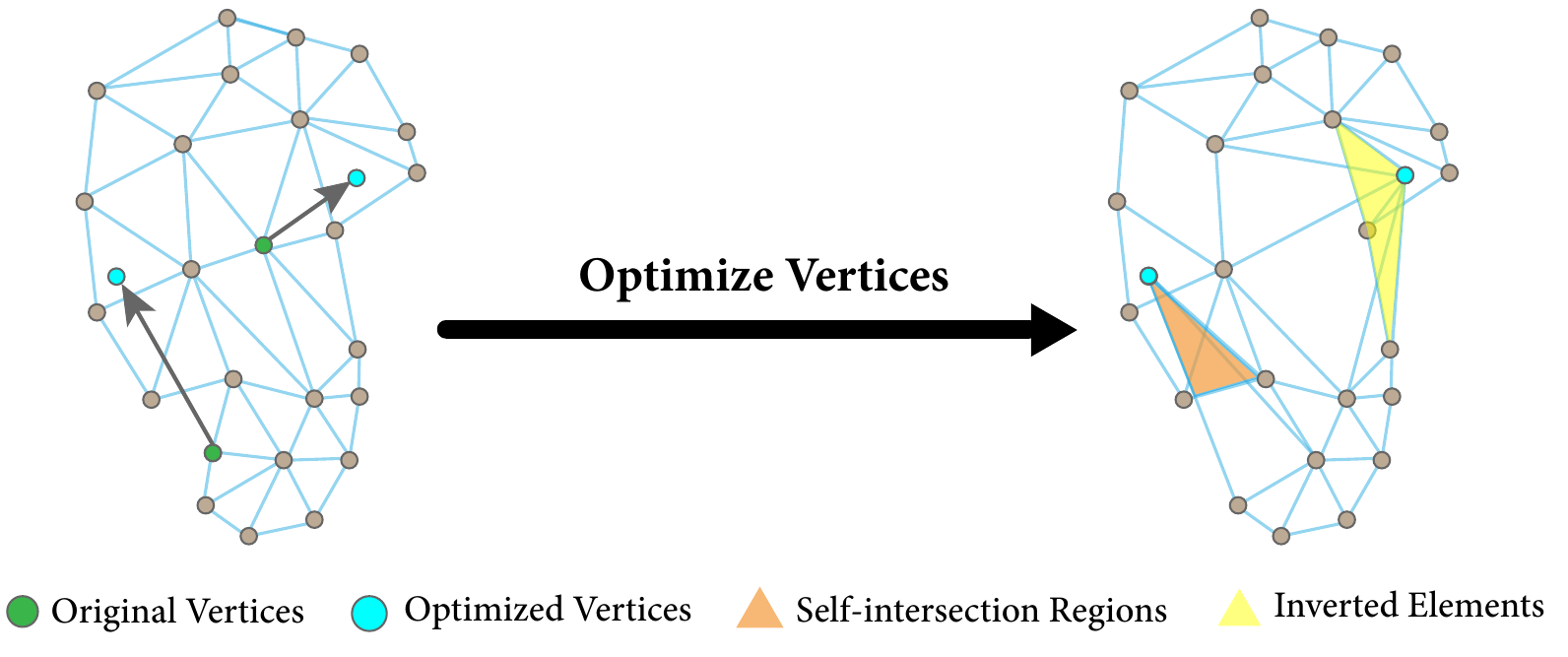}
    \caption{\textbf{Two causes leading to mesh anomalies: self intersection and element inversion.} }
    \label{fig:issues}
\end{figure}

One of our goals is to optimize the base mesh $\mathcal{M}$ while maintaining high-quality structure. As shown in Fig.~\ref{fig:abla} (W/o constraints), optimizing a tetrahedral mesh without any constraints leads to poor mesh quality. The main cause of this issue is that the vertices can freely move during the optimization of the tetrahedral mesh. Consequently, the tetrahedral mesh suffers from self-intersections and element inversion problems, resulting in tetrahedra overlaps as shown in Fig.~\ref{fig:issues}.

A naive solution is to introduce a loss function to constrain the mesh optimization process. Locally injective mapping~\cite{schuller2013locally} uses a barrier function to strictly prevent element inversion. However, the gradient explosion near the critical points of the barrier function makes it difficult to use directly in gradient descent-based optimization. Another approach is to apply an $L_1$-norm constraint on the signed volume of the optimized tetrahedra $L_{\text{sv}}$, making tetrahedra with negative signed volume sparse:
\begin{align}
\mathcal{L}_{\text{sv}} &= \frac{1}{K} \sum_{k=1}^K \max \left( -s(T_k), 0 \right),
\end{align}
where $s(T_k)$ is the signed volume of the tetrahedron $T_k$, $T^0_k$ is the initial tetrahedron. However, we found that these loss function-based constraints significantly impact the rendering quality and do not fully guarantee that the tetrahedral mesh remains a reasonable structure as shown in Sec.~\ref{sec:6.2}.

\paragraph{Orientation-preserving Homeomorphism as constraint.} Unlike previous methods that used regularization loss, we adopt the orientation-preserving homeomorphism as a constraint, implicitly restricting the feasible region of the vertex set \(\mathcal{V}\) of the tetrahedral mesh. The two key properties of this mapping are:

\begin{itemize}
    \item \textbf{Orientation-preserving} means that the Jacobian determinant of the mapping remains positive, which prevents element inversion during tetrahedral mesh optimization.
    \item \textbf{Homeomorphism} guarantees that the tetrahedral mesh after optimization does not produce self-intersecting boundaries.
\end{itemize}

During optimization, both the topology $\mathcal{T}$ and vertex positions $\mathcal{V}$ of the base mesh $\mathcal{M}$ are fixed. The orientation-preserving homeomorphism maps a vertex \(v = [x, y, z]\) to its actual position as follows:
\begin{equation}
v' = [x', y', z']^T = \mathcal{H}([x, y, z]^T).
\label{eq:hom}
\end{equation}
We use the mapped vertices \( \mathcal{V}' = \{ v_i' \} \) to reparameterize the 3D Gaussians. This orientation-preserving homeomorphism can be viewed as a constraint on the feasible region of the vertices, ensuring that a valid mesh is still generated given a topology \( \mathcal{T} \).

\paragraph{Real-NVP} Based on these observations, we implement the orientation-preserving homeomorphism $\mathcal{H}$ by a novel invertible network. Real-NVP~\cite{dinh2016density} appears to be a reasonable network architecture, as it is bijective and ensures that the Jacobian determinant of each coupling layer remains positive. Real-NVP splits the input $x\in \mathbb{R}^D$ into two components, \( x_{1:d} \) and \( x_{d+1:D} \). The output of a coupling layer $x^{\prime}\in \mathbb{R}^D$ follows the equations:

\begin{equation}
\begin{aligned}
x^{\prime}_{1:d} = x_{1:d},\quad x^{\prime}_{d+1:D} = x_{d+1:D} \cdot \exp(s(x_{1:d})) + t(x_{1:d}),
\end{aligned}
\end{equation}
where \( s: \mathbb{R}^d \to \mathbb{R}_+^{D-d} \) and \( t: \mathbb{R}^d \to \mathbb{R}^{D-d} \) are the scale and translation functions. We denote this map of the coupling layer as \( h: \mathbb{R}^D \to \mathbb{R}^D \). The Jacobian of this mapping is:
\begin{equation}
J_h = \begin{bmatrix}
I & 0  \\
\frac{\partial x^{\prime}_{d+1:D}}{\partial x^T_{1:d}} & \text{diag}[\exp(s(x_{1:d})]
\end{bmatrix}.
\end{equation}
This is a lower triangular matrix, so its determinant is the product of the diagonal elements, which is positive. Real-NVP stacks multiple coupling layers in an alternating pattern, such that the components that are left unchanged in one coupling layer are updated in the next coupling layer. However, when we split \( (x, y, z) \) into two components: \( (x, z) \) and \( y \), after mapping by a similar coupling layer, the Jacobian matrix is in the following form, which is not a triangular matrix, and its determinant does not remain positive.
\begin{equation}
J_h = \begin{bmatrix}
1 & \frac{\partial y'}{\partial x} & 0 \\
0 & \exp(s) & 0 \\
0 & \frac{\partial y'}{\partial z} & 1
\end{bmatrix}
\end{equation}

\paragraph{Orientation-preserving Networks} The above problem is caused by the fact that the input is split into discontinuous parts, resulting in the Jacobian matrix being neither an upper triangular matrix nor a lower triangular matrix. To solve this problem, we propose a permutation strategy. As shown in Fig.~\ref{fig:inn}, we multiply the input by a permutation matrix \( P \), ensuring that the input is always divided into contiguous two parts. For example, when we need to keep the \( x \) and \( z \) components unchanged and transform the \( y \) component, we apply the permutation matrix \( P \) to the input and output. Therefore, the permuted transformation mapping \( \phi: \mathbb{R}^3 \to \mathbb{R}^3 \) is:
\begin{equation}
\phi\left( [x , y , z]^T \right) = P^{-1} \left( h \left( P [x , y , z]^T \right) \right), \quad P = \begin{bmatrix} 0 & 0 & 1 \\ 1 & 0 & 0 \\ 0 & 1 & 0 \end{bmatrix}.
\end{equation}

Thus, the Jacobian matrix is: $J_{\phi} = P J_h P^{-1}$, where \( J_h \) is the Jacobian matrix of the coupling layer \( h \). Since \( J_h \) is an upper triangular matrix, and its diagonal elements are positive, it satisfies \( \det(J_h) > 0 \). As \( \det(P) > 0 \), it follows that \( \det(J_{\phi}) > 0 \).

\subsection{Training}
\label{sec:training}

\paragraph{Initialization} We use two different initialization strategies. In the first strategy, we reconstruct the surface using NeuS2~\cite{wang2023neus2}, and then generate the tetrahedral mesh from the reconstructed triangle mesh using fTetWild~\cite{hu2020fast}. The second is to generate a uniform tetrahedral grid within the bounding box of the scene. We conduct experiments on different initialization strategies in Sec.~\ref{sec:6.2}, which show that even with poorly initialized tetrahedral mesh, we can still achieve high-quality rendering results.


\paragraph{Adaptive Control of Primitives}
As detailed in Sec.~\ref{sec:conformal}, our primary mechanism for adapting geometric detail for rendering is the hierarchical implicit subdivision strategy. This approach inherently avoids direct modification of the base mesh $\mathcal{M}$ topology. To manage the set of active rendering primitives in $\mathcal{M}_{r}$ during optimization, we introduce an adaptive masking mechanism. Specifically, rendering tetrahedra whose opacity $o_k$ falls below a predefined threshold $\epsilon$ are masked. These masked tetrahedra no longer contribute to the rendering process. This strategy contrasts with directly deleting tetrahedra from the fixed-topology base mesh, which could introduce problematic structural changes or compromise conformality. 

\paragraph{Optimization} 

Our model is trained by minimizing a composite loss function:
\begin{equation}
\mathcal{L}=\mathcal{L}_{render} + \lambda_3 \mathcal{L}_{\text {mask}} +\lambda_4 \mathcal{L}_{\text{quality}} .
\end{equation} 
The loss used for supervision of the RGB signal: $  \mathcal{L}_{render} = \lambda_1 \mathcal{L}_{\mathrm{1}} + \lambda_2 \mathcal{L}_{\mathrm{SSIM}} $ follow 3D Gaussian splatting~\cite{kerbl3Dgaussians}. \( \mathcal{L}_{mask} \) is designed to prevent artifacts at the object boundaries:
\[
\mathcal{L}_{\textrm{mask}} = \| M - \hat{M} \|_1,
\]
where \( M \) represents the rendered opacity, and \( \hat{M} \) represents the ground truth opacity.

To specifically enhance the geometric quality of the base tetrahedral mesh, we introduce $\mathcal{L}_{\text{quality}}$~\cite{lo1997optimization}:

\begin{equation}
    \mathcal{L}_{\text{quality}} = \frac{1}{K} \sum_{k=1}^K max(r - Q,0),\quad Q =  \frac{C_0V}{S_{\textup{rms}}},
    \label{eq:quality}
\end{equation}
where $V$ represents the tetrahedron's volume, $C_0$ is a normalization constant to ensure $Q=1$ for an equilateral tetrahedron, and $S_{rms}$ denotes the root-mean-square of the six edge lengths of the tetrahedron. This loss
essentially constrains the quality of the tetrahedron to be greater than $r$. 
\section{Experiments}
We first give the implementation details. The contribution analysis of proposed components is given in Sec.~\ref{sec:6.2}. Evaluations of our proposed representation with previous state-of-the-art non-editable and editable approaches are presented in Sec.~\ref{sec:performance}. We also show some applications of our proposed representation in Sec.~\ref{sec:4.3}.

\paragraph{Implementation Details}
We set \( \lambda_1 = 0.8 \), \( \lambda_2 = 0.2 \), \( \lambda_3 = 0.5 \), and \( \lambda_4 =10.0 \) for training the model. We use the ReLU function to prevent the weights of the vertices from being negative, and apply the sigmoid function to ensure that the opacity stays within the range of $[0, 1)$. The gradient threshold $\delta$ for tetrahedron splitting is set to 0.0002, the same value used in 3DGS. We set the pruning threshold \( \epsilon = 0.05 \) to mask low-opacity tetrahedra. The invertible neural network consists of 3 blocks, with each axis chosen in order. A 2D hash encoding is used with a size of \( 2^{19} \) and a maximum resolution of \( 1024^2 \). The MLP has 2 hidden layers, each with 128 units. Additionally, we reconstruct the surface with NeuS2~\cite{wang2023neus2} and generate the initial tetrahedral mesh through fTetWild~\cite{hu2020fast}. Furthermore, we limit the maximum subdivision depth for each tetrahedron to 5 levels. We train each model with 30,000 iterations, which takes approximately 1 hour. All experiments are conducted on a single NVIDIA RTX4090 GPU.

\subsection{Ablation Study}
\label{sec:6.2}
In this section, we conduct two main ablation studies to address the following questions: 
\begin{itemize} 
\item Q1: How much does our orientation-preserving homeomorphism and $\mathcal{L}_{\text{quality}}$ affect rendering quality and mesh quality? 
\item Q2: Does an initially low-quality mesh degrade our reconstruction quality? 
\end{itemize}

\begin{table}[tbp]
\caption{\textbf{Quantitative results with different constraints.} Mesh quality is measured by the Aspect Ratio Gamma.}
\centering
\resizebox{\columnwidth}{!}{%
  \renewcommand{\arraystretch}{1.2}
  \setlength{\tabcolsep}{4pt} 
  \begin{tabular}{lccc|ccc}
    \toprule
                    & \multicolumn{3}{c|}{\textbf{\textit{Rendering Quality}}} & \multicolumn{3}{c}{\textbf{\textit{Mesh Quality}}} \\ 
                    & \textbf{PSNR ↑} & \textbf{SSIM ↑} & \textbf{LPIPS ↓} & \textbf{Quality ↑} & \textbf{\# Inverted ↓} &   \\ \hline 

(a) W/o Constraints  &  34.89 & 0.978 & 0.0210 & $0.159\pm0.179$ & $\sim 83.2k$ & \\ 
(b) W/ $\mathcal{L}_{sv}$ & 33.59 & 0.975 & 0.0294 & $0.144\pm0.187$ & $\sim 53.6k$ & \\ 
(c) W/ $\mathcal{H}$ & 34.64  &  0.974 & 0.0220 & $0.471\pm0.259$ & $\sim 1.3k$ & \\ 
(d) W/ $\mathcal{H}$ + $\mathcal{L}_{quality}$  & 33.64  & 0.974  & 0.0455 & $0.816\pm0.058$ & 0 & \\ 
(e) W/o Optimization         & 32.78           & 0.971        & 0.0545        & $0.797\pm0.111$        & 0 & \\ 
\bottomrule
\end{tabular}
}

\label{tab:constraints}
\end{table}

\begin{figure*}[htbp]
    \centering
    \includegraphics[width=\linewidth]{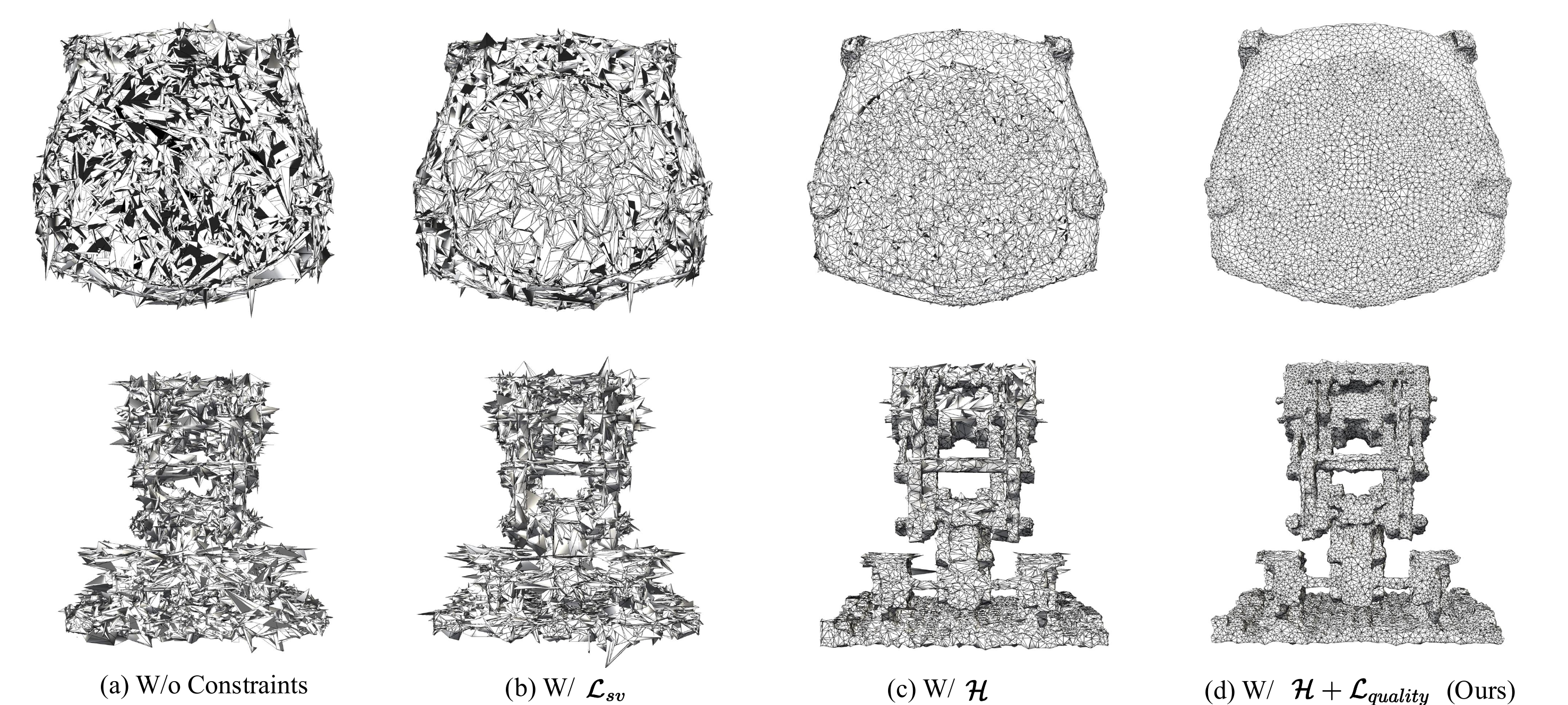}
    \caption{\textbf{Visualization of tetrahedral mesh cross-sections optimized with different constraints.} }
    \label{fig:abla}
\end{figure*}

\paragraph{Q1: Effect of Orientation-preserving Homeomorphism and $\mathcal{L}_{\textrm{quality}}$.} We conduct this experiment on the \textit{Lego} and  \textit{Chair} scenes of \textit{NeRF synthetic} dataset, with the mesh initialized by NeuS2 as input. To avoid the influence of number of primitives on the experiment, we remove the module of adaptive control of primitives,  fixing the number of tetrahedra. We regularize the training of the tetrahedral mesh using four different methods: (a) without explicit constraints, (b) with the signed volume loss $ \mathcal{L}_{\text{sv}} $, (c) with our orientation-preserving homeomorphism $\mathcal{H}$ as a constraint, and (d) with both $\mathcal{H}$ and our mesh quality loss $\mathcal{L}_{\text{quality}}$ applied together. Furthermore, we provide a (e) baseline where the tetrahedral mesh vertices are not optimized, and only other rendering attributes are refined.

Tab.~\ref{tab:constraints} and Fig.~\ref{fig:abla} present a comparison of reconstruction quality and geometric quality for these methods. Optimizing the mesh directly (without constraints) and optimizing with our homeomorphism $\mathcal{H}$ achieve comparable rendering quality. However, employing our $\mathcal{H}$ results in significantly fewer inverted elements and higher overall geometric quality. In contrast, applying the $\mathcal{L}_{\text{sv}}$ loss has a limited effect on reducing the number of inverted elements and noticeably degrades rendering quality. This suggests that our approach of defining the feasible optimization space for vertices via an orientation-preserving homeomorphism  is less susceptible to converging to poor local minima compared to methods based solely on regularization penalties. The geometric quality regularization term $\mathcal{L}_{\text{quality}}$ leads to a slight decrease in rendering quality compared to using $\mathcal{H}$ alone, but it yields a substantial improvement in the geometric quality of the mesh elements.  

\begin{figure*}[htbp]
    \centering
    \includegraphics[width=1.0\linewidth]{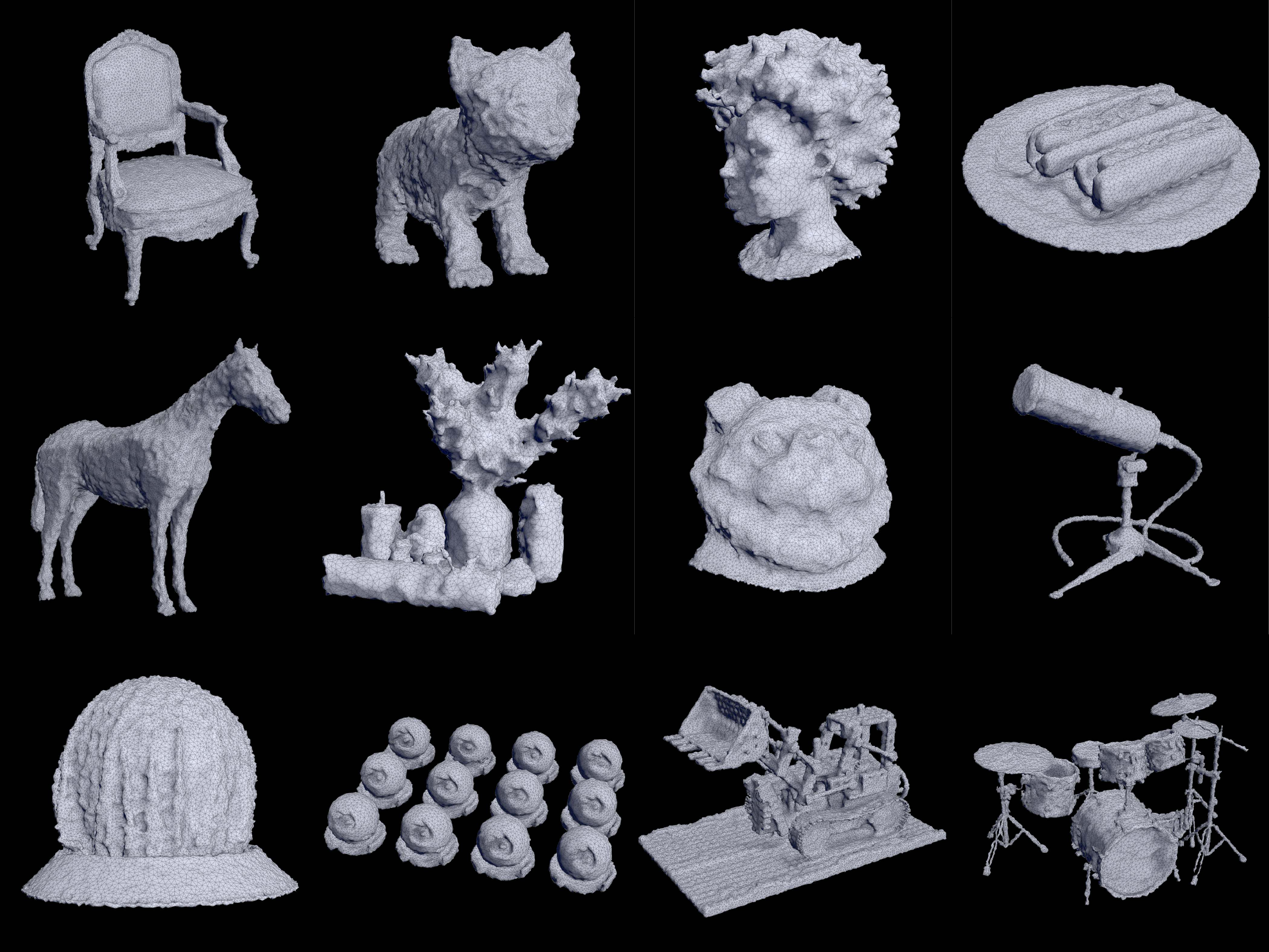}
    \caption{\textbf{ Gallery of mesh results on \textit{Shelly} dataset and \textit{NeRF Synthetic} dataset.}}
    \label{fig:more_mesh}
\end{figure*}

We present more visual results of the optimized mesh in Fig.~\ref{fig:more_mesh} and numerical results in Tab.~\ref{tab:more_tab3} and Tab.~\ref{tab:more_tab3}. The reported mesh quality metrics include AR (Aspect Ratio), defined as $3r_k/\rho_k$ (where $r_k$ is the inradius, and $\rho_k$ is the circumradius of tetrahedron $T_k$). ARG (Aspect Ratio Gamma) is consistent with the quality metric $Q$ in Eq.~\ref{eq:quality}. Lastly, AVR (Adjacent cells Volume Ratio) measures the ratio of the largest to smallest cell volume within the 1-ring neighborhood of a tetrahedron. Notably, after constrain with our orientation-preserving homeomorphism $\mathcal{H}$ and $\mathcal{L}_{\text{quality}}$ term, the geometric quality of our final mesh often surpasses that of the initial mesh.

\begin{table}[tbp]
\caption{\textbf{Quantitative results for different initializations.}}
\centering

\renewcommand{\arraystretch}{1.1} 
\setlength{\tabcolsep}{1pt} 
\small 

\begin{tabular}{l c c|c c|c c}
    \toprule
            & \multicolumn{2}{c|}{\textbf{\textit{lego}}}   & \multicolumn{2}{c|}{\textbf{\textit{hotdog}}} & \multicolumn{2}{c}{\textbf{\textit{mic}}} \\ \cline{2-7} 
            & \textbf{PSNR ↑} & \textbf{SSIM ↑} & \textbf{PSNR ↑} & \textbf{SSIM ↑} & \textbf{PSNR ↑} & \textbf{SSIM ↑} \\ \hline
W/o Rotation     & 33.27   & 0.968 & 36.14 & 0.975 & 36.06 &  0.986 \\
W/ Uniform       & 34.98            & 0.980           & 36.90            & 0.983           & 37.46            & 0.991           \\
W/ NeuS2        & 35.50            & 0.983           & 37.80            & 0.986           & 37.64            & 0.993           \\ 
    \bottomrule
\end{tabular}

\label{tab:abal2}
\end{table}

\paragraph{Q2: Ablations on Initial Mesh.}  We conduct this experiment on the \textit{NeRF synthetic} dataset: lego, hotdog, mic. Initially, we start with a uniform tetrahedral mesh as described in Sec.~\ref{sec:training}. This tetrahedral mesh is then used as input into our method for training. 
Tab.~\ref{tab:abal2} shows the final reconstruction results under different settings compared to those obtained with NeuS2 initialization. Notably, ``W/o Rotation'' means the removal of the optimizable rotation parameters in Eq.~\ref{eq:final_covariance}, which leads to a significant degradation in reconstruction quality. This performance drop is likely because the per-vertex weights are then solely responsible for determining both the shape and effective orientation of the 3D Gaussians, thereby limiting their expressive ability. Conversely, when the optimizable rotation parameters are included, the ``W/ Uniform'' initialization setting still produces reconstruction results comparable to those achieved with the NeuS2 initialization (``W/ NeuS2 '').

\subsection{Comparison}
\label{sec:performance}
We conduct experiments to compare the performance of rendering on static scenes. The results show that, despite the need to maintain a structured geometry during optimization, we are still able to achieve comparable results to that of unstructured radiance field representation like 3DGS~\cite{kerbl3Dgaussians}.
 
\begin{table}[tbp]
\caption{\textbf{Quantitative results on \textit{NeRF Synthetic} and \textit{Shelly} dataset.}}
\centering

\resizebox{\columnwidth}{!}{%
  \renewcommand{\arraystretch}{1.2}
  \setlength{\tabcolsep}{4pt} 
  \begin{tabular}{lccc|cccc}
    \toprule
            & \multicolumn{3}{c|}{\textbf{\textit{NeRF Synthetic}}} & \multicolumn{4}{c}{\textbf{\textit{Shelly}}} \\ \cline{2-8} 
            & \textbf{PSNR ↑} & \textbf{SSIM ↑} & \textbf{LPIPS ↓} & \textbf{PSNR ↑} & \textbf{SSIM ↑} & \textbf{LPIPS ↓} &  \\ \hline
NeRF          & 31.00            & 0.947           & 0.081           & 31.28            & 0.893           & 0.157           &  \\
Adaptive Shells & 31.84            & 0.957           & 0.056           & 36.02            & 0.954           & 0.079           &  \\
3DGS          & \cellcolor{red!25}\textbf{33.78}            & \cellcolor{red!25}\textbf{0.969}           & \cellcolor{orange!25}0.030           & \cellcolor{orange!25}39.61            & \cellcolor{orange!25}0.961           & \cellcolor{yellow!25}0.064           &  \\
2DGS          & \cellcolor{yellow!25}32.92            & \cellcolor{yellow!25}0.966           & 0.036              & 34.62              & 0.928              & 0.098              &  \\
Mani-GS       & 32.58               & 0.961              & \cellcolor{yellow!25}0.034              & --               & --              & --              &  \\
SuGaR         & 30.94            & 0.954           & 0.040           & \cellcolor{yellow!25}36.43            & \cellcolor{yellow!25}0.956           & \cellcolor{orange!25}0.057           &  \\
Ours          & \cellcolor{orange!25}33.53            & \cellcolor{orange!25}0.968           & \cellcolor{red!25}\textbf{0.019}           & \cellcolor{red!25}\textbf{39.76}            & \cellcolor{red!25}\textbf{0.967}           & \cellcolor{red!25}\textbf{0.052}           &  \\ 
    \bottomrule
  \end{tabular}
}

\label{tab:tab1}
\end{table}

\begin{table}[tbp]
\caption{\textbf{Quantitative results on \textit{Mip-NeRF360} dataset.} Results of other methods are borrowed from Radiant Foam's paper.}
\centering

\resizebox{\columnwidth}{!}{%
  \renewcommand{\arraystretch}{1.2}
  \setlength{\tabcolsep}{4pt} 
  \begin{tabular}{lccc|cccc}
    \toprule
            & \multicolumn{3}{c|}{\textbf{\textit{Outdoor Scene}}} & \multicolumn{4}{c}{\textbf{\textit{Indoor Scene}}} \\ \cline{2-8} 
            & \textbf{PSNR ↑} & \textbf{SSIM ↑} & \textbf{LPIPS ↓} & \textbf{PSNR ↑} & \textbf{SSIM ↑} & \textbf{LPIPS ↓} &  \\ \hline
3DGS          & {\cellcolor{orange!25}26.40}            & {\cellcolor{orange!25}0.807}           & {\cellcolor{yellow!25}0.197}           & 30.38            & {\cellcolor{yellow!25}0.920}           & 0.220           &  \\
Mip‑Splatting & {\cellcolor{red!25}\textbf{26.81}}      & {\cellcolor{red!25}\textbf{0.817}}     & {\cellcolor{red!25}\textbf{0.170}}     & {\cellcolor{orange!25}31.19}            & {\cellcolor{red!25}\textbf{0.933}}     & 0.223           &  \\
MipNerf360    & {\cellcolor{yellow!25}25.92}            & 0.747                                  & 0.243                                  & {\cellcolor{red!25}\textbf{31.72}}      & 0.915                                  & {\cellcolor{orange!25}0.180}           &  \\
Radiant Foam  & 25.42                                  & 0.737                                  & 0.253                                  & 30.75                                   & 0.908                                  & {\cellcolor{red!25}\textbf{0.170}}      &  \\
Ours          & 25.71                                  & {\cellcolor{yellow!25}0.774}           & {\cellcolor{orange!25}0.193}           & {\cellcolor{yellow!25}30.83}            & {\cellcolor{orange!25}0.921}           & {\cellcolor{yellow!25}0.197}           &  \\  
    \bottomrule
  \end{tabular}
}

\label{tab:tab2}
\end{table}

\begin{figure*}[htbp]
    \centering
    \includegraphics[width=1.0\linewidth]{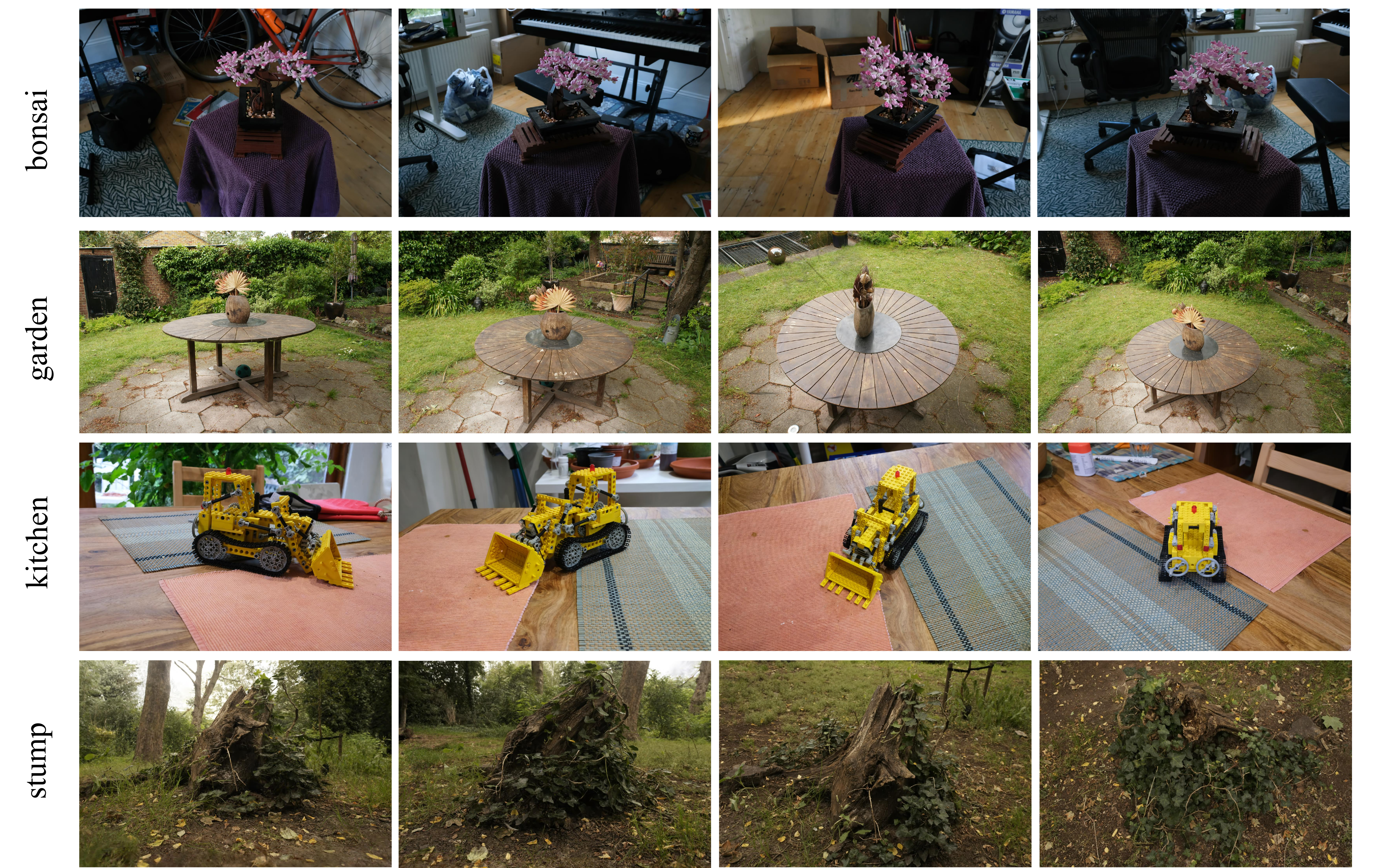}
    \caption{\textbf{ Gallery of Rendering Results on \textit{MipNeRF 360} dataset.}}
    \label{fig:more_lod}
\end{figure*}

\begin{figure*}[htbp]
    \centering
    \includegraphics[width=1.0\linewidth]{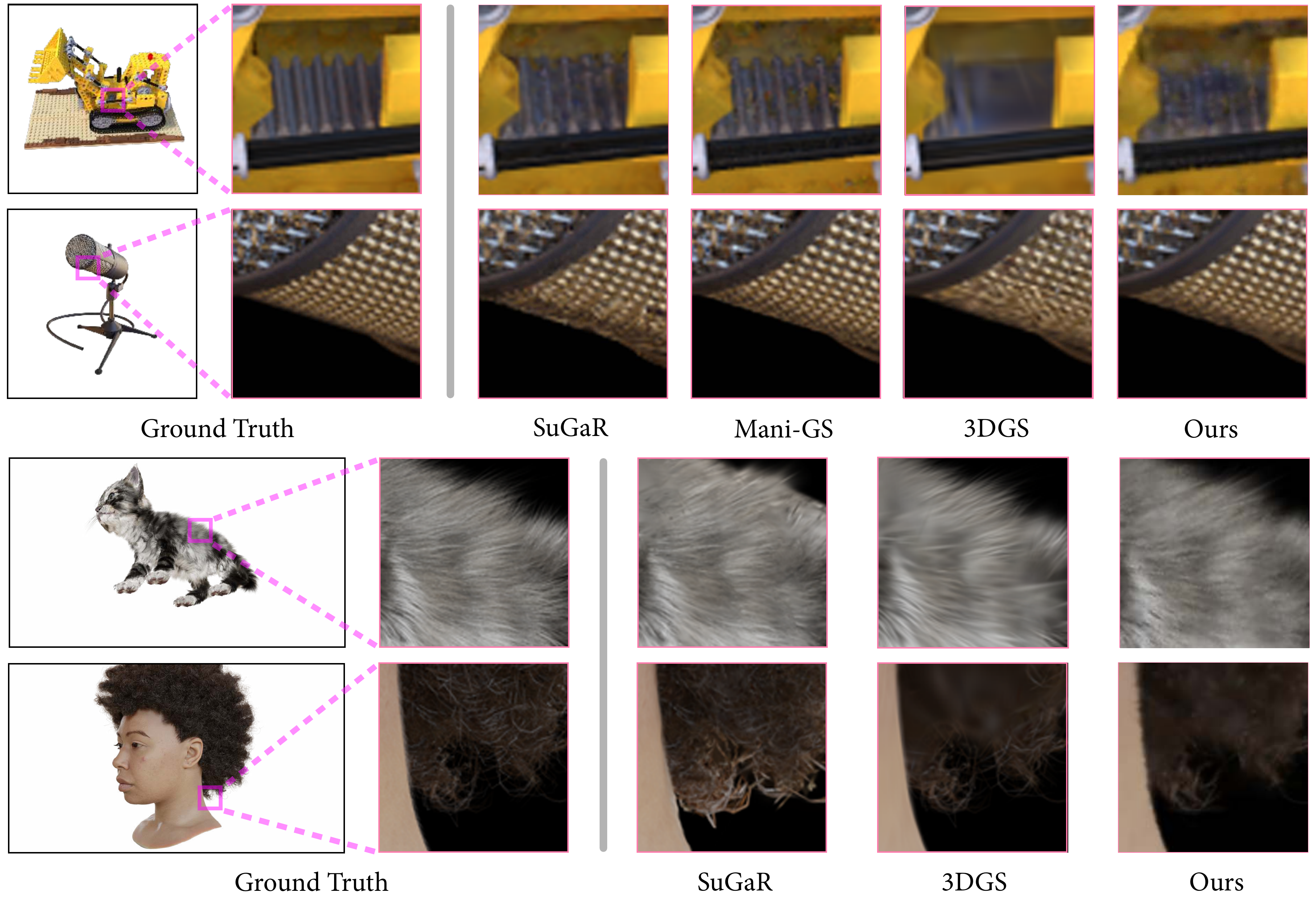}
    \caption{\textbf{Our results on the test-views of \textit{Shelly} dataset and \textit{NeRF Synthetic} dataset.}}
    \label{fig:comparison1}
\end{figure*}

We employ two datasets for evaluation: \textit{NeRF Synthetic} dataset~\cite{mildenhall2021nerf}, containing eight scenes and \textit{Shelly} dataset~\cite{wang2023adaptive}, containing six scenes. We conduct comparisons with implicit, point-based, and mesh-based methods to evaluate the performance of our approach, including NeRF~\cite{mildenhall2021nerf}, 3DGS~\cite{kerbl3Dgaussians}, 2DGS~\cite{Huang2DGS2024}, Adaptive Shells~\cite{wang2023adaptive} and Mani-GS~\cite{gao2024mani}. We use the surface reconstructed by NeuS2 as input to Mani-GS for a fair comparison. We use the same training process as 3DGS.  The rendering quality is evaluated using a set of established metrics: PSNR, SSIM~\cite{wang2004image}, and LPIPS~\cite{zhang2018unreasonable}. In Tab.~\ref{tab:tab1}, we conduct a quantitative comparison of rendering quality. Our approach shows competitive results across all mesh-based and point-based methods according to three evaluation metrics. Moreover, our approach achieves qualitatively better reconstructions, with fewer artifacts and more detailed results, as shown in Fig.~\ref{fig:comparison1}.

Furthermore, to demonstrate the effectiveness of our method in real-world unbounded scenes, we conducted experiments on the MipNeRF 360 dataset~\cite{barron2022mip}. For initialization in these scenes, we utilize a mesh of the centrally located region, initially reconstructed by Gaussian Opacity Fields~\cite{yu2024gaussian}. This mesh is then tetrahedralized using fTetWild~\cite{hu2020fast} to serve as the input tetrahedral mesh for our method. Our StructuredField representation is employed to reconstruct this central region of interest, while the surrounding background is modeled using standard 3D Gaussian Splatting ~\cite{kerbl3Dgaussians}. Tab.~\ref{tab:tab2} presents a numerical comparison of our method against several state-of-the-art approaches, including 3DGS~\cite{kerbl3Dgaussians}, Mip-Splatting~\cite{yu2004mesh}, MipNeRF360~\cite{barron2022mip}, and RadiantFoam~\cite{govindarajan2025radiant}. Our method achieves reconstruction results comparable to those of 3DGS and surpasses RadiantFoam, another approach that also utilizes a deformable tetrahedral grid as its geometric primitive for scene representation. Additional visual results from our method on these scenes are provided in Fig.~\ref{fig:more_lod}.

\subsection{Applications}
\label{sec:4.3}

\begin{figure*}[htbp]
    \centering
    \includegraphics[width=1.0\linewidth]{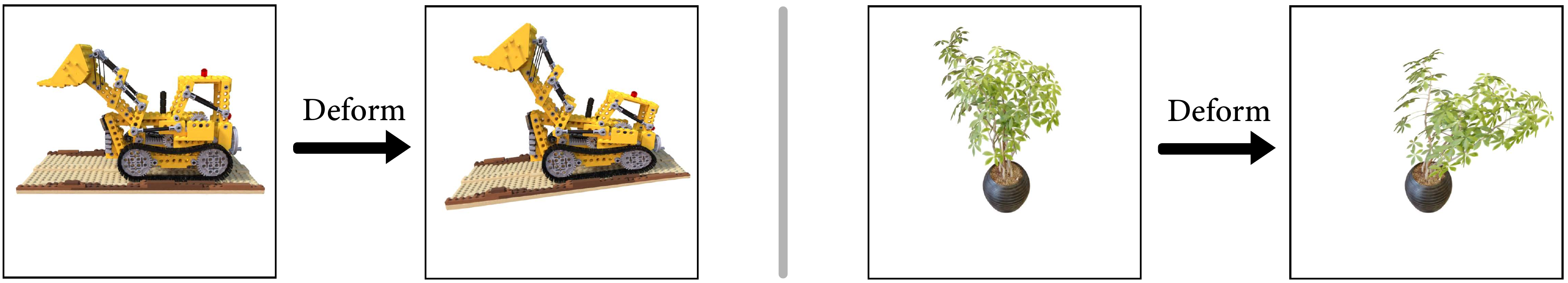}
    \caption{\textbf{ Gallery of deformation results.}  Videos are provided in the supplementary materials for further visualization.}
    \label{fig:deformation}
\end{figure*}

Our method directly constructs an explicit tetrahedral mesh $\mathcal{M}$ with good quality and achieves good rendering quality at the same time, which proves highly valuable for a variety of downstream applications. With tetrahedral mesh, we can perform physics simulations, animations, editing, and other operations seamlessly. With our proposed StructuredField, we can easily update the mesh by simply tracking the movement or displacement of each vertex. This allows us to reparameterize the Gaussians based on the new positions of vertices, enabling the rendering of deformed scenes.

Methods like Mani-GS and GaMes are hybrid representations that fix mesh vertex positions and establish a relationship between triangle meshes and 3D Gaussians. In contrast, our method offers a unified, structured representation where the vertex positions are optimized during the reconstruction process. Since the final meshes produced by our method differ from those in hybrid methods, it is difficult to establish a fair comparison under the same deformation.

For many applications such as physical simulation and deformation, the tetrahedral mesh representation in our StructuredField has inherent advantages. Unlike triangle meshes, which only represent the surface of an object, tetrahedral meshes have an internal structure, making it more suitable for physically accurate simulations. On the other hand, thanks to our carefully designed reparameterization strategy, the primitives in our representation are always confined within the corresponding tetrahedra. This ensures that our representation accurately reflects the mesh deformation. In contrast, previous methods using triangle meshes may suffer from artifacts during deformation, as the 3D Gaussians could shift away from the triangular surfaces, leading to potential inconsistencies between 3D Gaussians and triangle meshes.

\begin{figure*}[htbp]
    \centering
    \includegraphics[width=1.0\linewidth]{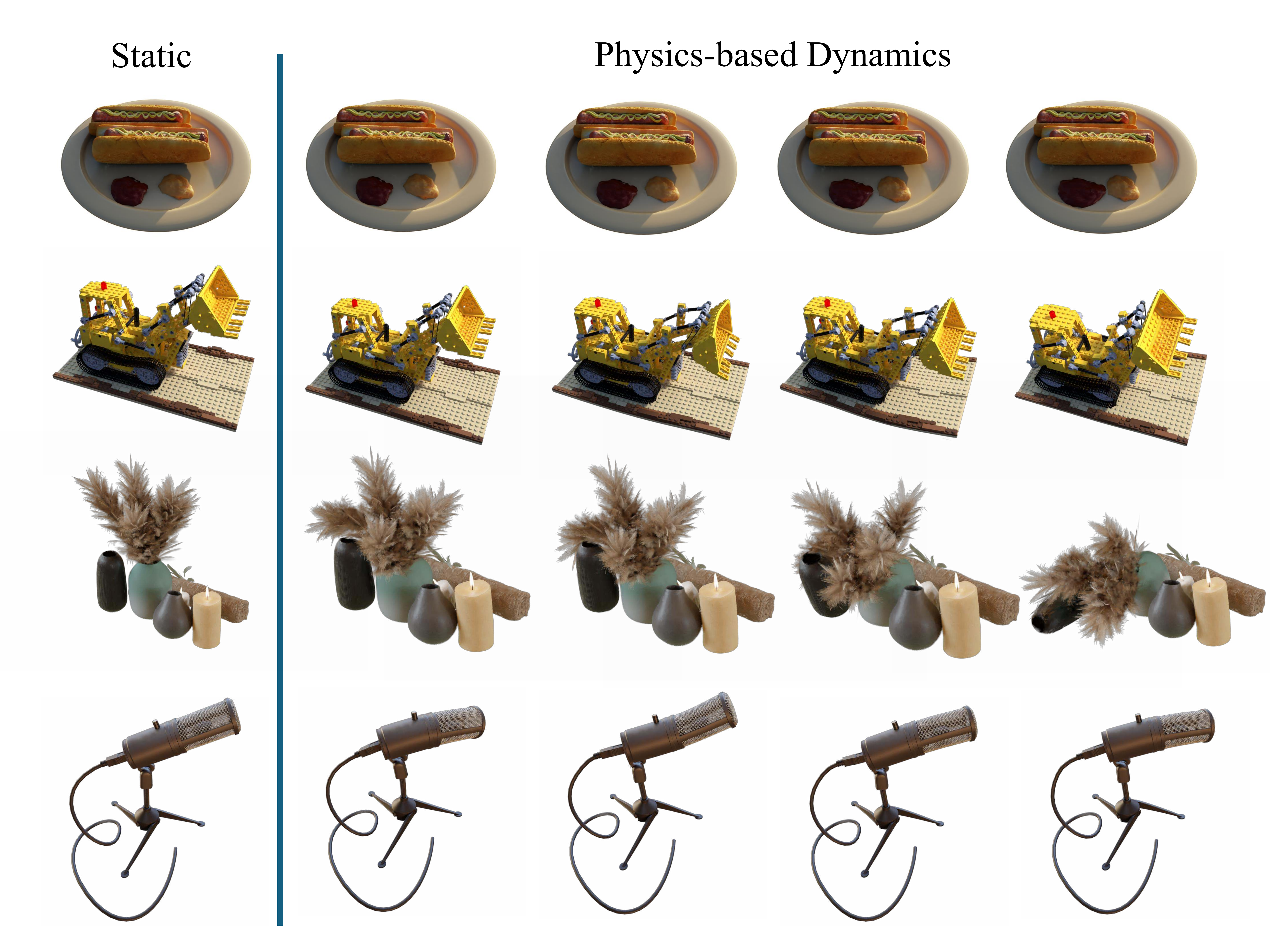}
    \caption{\textbf{Gallery of physical simulation results.} The original shape is shown in the leftmost column, with the corresponding physical simulations displayed in the right column. Videos are provided in the supplementary materials for further visualization.}
    \label{fig:simulation}
\end{figure*}

\paragraph{Physical Simulation}

After constructing the base tetrahedral mesh $\mathcal{M}$, we use it for physical simulation. This process yields a sequence of vertex positions defining its dynamic state over time, denoted as $\mathcal{M}^{k}$. The motion of any implicitly defined child tetrahedra is then consistently derived from their parent elements within $\mathcal{M}^{k}$. This is achieved by applying the previously optimized barycentric coordinates of the subdivision control points to the vertices of the dynamically deformed parent tetrahedra. This process yields the dynamic rendering mesh, $\mathcal{M}_r^{k}$.

To simulate elastic deformations specifically, we treat the base mesh $\mathcal{M}$ as a mass-spring system. In this system, each vertex is modeled as a mass point, and the edges of the tetrahedra serve as springs connecting these points. We implement the XPBD algorithm~\cite{XPBD} using the Taichi programming language~\cite{Taichi}. Other types of simulation effects or complementary demonstrations presented are realized in Houdini~\cite{Houdini}. Fig.~\ref{fig:simulation} shows results from our physics simulation approach; please refer to the supplementary video for dynamic motion.

\paragraph{Deformation}
We utilize lattice deformation as a mechanism to drive the deformation of the tetrahedral mesh. As the mesh vertices undergo deformation, we reparameterize the associated Gaussian primitives based on the updated positions of the mesh vertices. As demonstrated in Fig.~\ref{fig:deformation}, even under large-scale deformations, our representation still produces reasonable rendering results.

\begin{figure*}[htbp]
    \centering
    \includegraphics[width=1.0\linewidth]{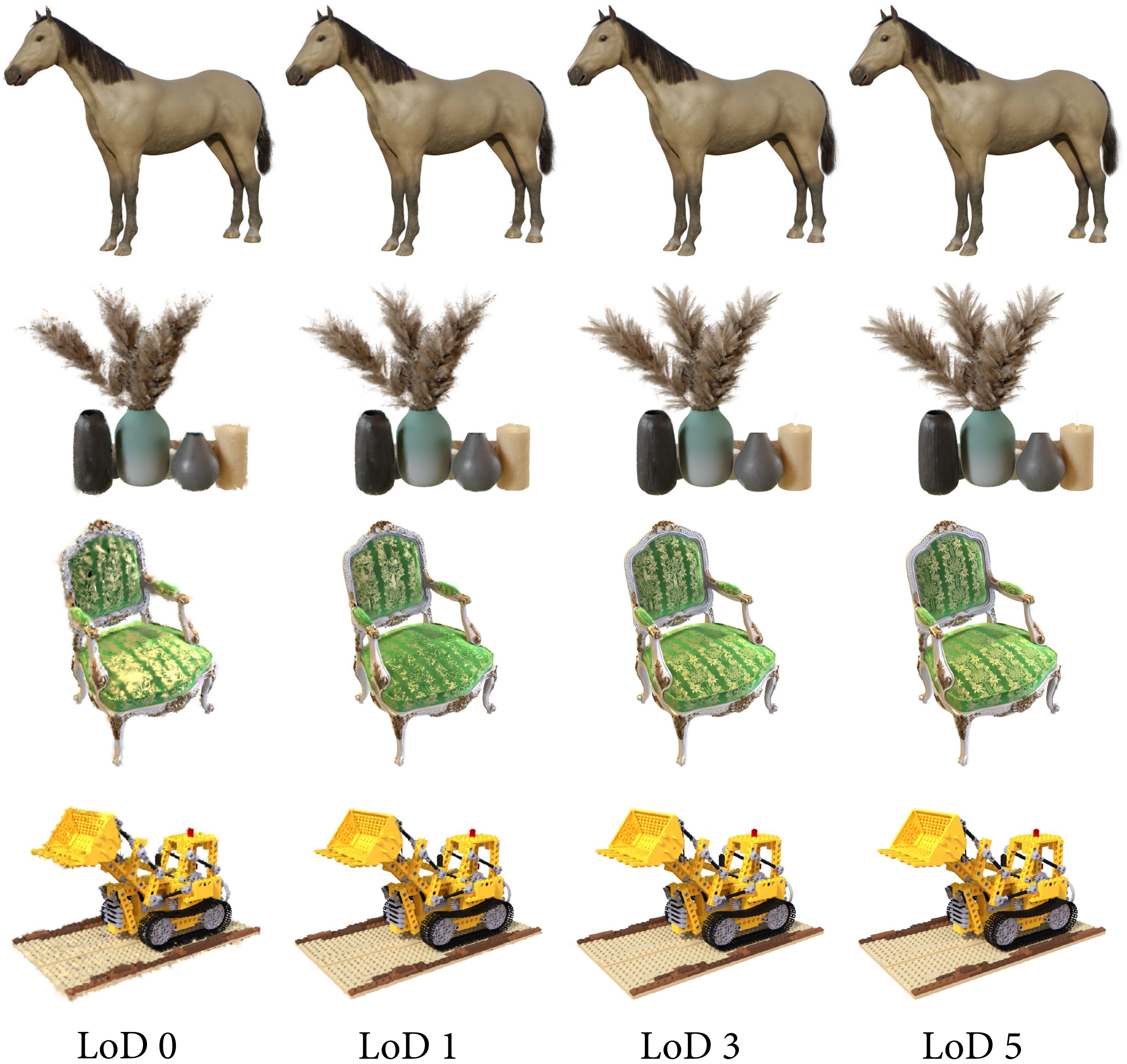}
     \caption{\textbf{Demonstration of our method's LoD capabilities on \textit{Shelly} Dataset and \textit{NeRF Synthetic}  Dataset.}}
    \label{fig:lod}
\end{figure*}

\paragraph{Level-of-Detail} The hierarchical nature of our implicit subdivision naturally facilitates an efficient Level of Detail (LOD) rendering strategy. This allows the complexity of the rendered scene to be dynamically adjusted, balancing visual fidelity with computational performance. Starting from the rendering mesh $\mathcal{M}_r$ , coarser LODs are generated by progressively replacing sets of four child tetrahedra with their respective parent tetrahedron. For attributes defined per tetrahedron, the opacity of the parent tetrahedron  is taken as the maximum of its four children's opacities, and its rotation quaternion  is determined by averaging the rotation quaternions of its children. Fig.~\ref{fig:lod} demonstrates the effects of our LOD approach.

\begin{table*}[tbp]
\caption{\textbf{Quality metrics of initial mesh and optimized mesh on \textit{NeRFSynthetic} Dataset.}}
\centering
\resizebox{\textwidth}{!}{%
  \renewcommand{\arraystretch}{1.2}
  \setlength{\tabcolsep}{4pt}
  \begin{tabular}{l|ccc|ccc|ccc|ccc}
    \toprule
     & \multicolumn{3}{c|}{\textbf{lego}}
     & \multicolumn{3}{c|}{\textbf{mic}}
     & \multicolumn{3}{c|}{\textbf{ship}}
     & \multicolumn{3}{c}{\textbf{chair}} \\
    \cline{2-13}
      & \textbf{AR↑}
      & \textbf{ARG↑}
      & \textbf{AVR↓}
      & \textbf{AR↑}
      & \textbf{ARG↑}
      & \textbf{AVR↓}
      & \textbf{AR↑}
      & \textbf{ARG↑}
      & \textbf{AVR↓}
      & \textbf{AR↑}
      & \textbf{ARG↑}
      & \textbf{AVR↓} \\
    \midrule
    Initial Mesh
      & $0.853\pm0.086$ & $0.829\pm0.091$ & $1.234\pm0.228$
      & $0.826\pm0.105$ & $0.800\pm0.109$ & $1.274\pm0.297$
      & $0.823\pm0.118$ & $0.798\pm0.121$ & $1.306\pm0.358$
      & $0.848\pm0.097$ & $0.824\pm0.101$ & $1.255\pm0.285$ \\
    Ours Mesh
      & $\textbf{0.855}\pm\textbf{0.053}$ 
      & $\textbf{0.829}\pm\textbf{0.056}$ 
      & $\textbf{1.229}\pm\textbf{0.202}$
      & $\textbf{0.849}\pm\textbf{0.051}$ 
      & $\textbf{0.821}\pm\textbf{0.054}$ 
      & $\textbf{1.236}\pm\textbf{0.208}$
      & $\textbf{0.849}\pm\textbf{0.054}$ 
      & $\textbf{0.822}\pm\textbf{0.056}$ 
      & $\textbf{1.254}\pm\textbf{0.226}$
      & $\textbf{0.854}\pm\textbf{0.053}$ 
      & $\textbf{0.827}\pm\textbf{0.056}$ 
      & $\textbf{1.238}\pm\textbf{0.213}$ \\
    \bottomrule
  \end{tabular}%
}
\label{tab:more_tab3}
\end{table*}

\section{Conclusion}

We introduced StructuredField, a novel structured 3D representation that unifies high-fidelity rendering and structured geometry. Our main contribution is using reparameterization to make tetrahedral mesh rendering differentiable. An orientation-preserving homeomorphism has been proposed to ensure the mesh quality during optimization. Extensive experiments have been conducted to verify the effectiveness of StructuredField in the aspects of rendering quality, physics simulations, and deformation modeling.

\paragraph{Limitations} Although our proposed StructuredField representation can simultaneously recover high-quality rendering and structured geometry from multi-view images, there are still some limitations: First, during the training process, vertices need to pass through the neural network, which leads to a longer training time than unstructured radiance field representations. In addition, since we need to maintain structured geometry during the reconstruction process, our solution space is more restricted than unstructured radiance field representations, so a larger number of primitives are required to represent the scene to achieve similar rendering effects.


\begin{acks}
This research was supported by the National Natural Science Foundation of China (No.62122071, No.62272433), the Youth Innovation Promotion Association CAS (No. 2018495) and the Fundamental Research Funds for the Central Universities (No. WK3470000021).
\end{acks}

\bibliographystyle{ACM-Reference-Format}
\bibliography{reference}

\end{document}